\documentstyle[epsfig,12pt]{article}
\begin{document}

{\bf
\title "The angular velocity of the apsidal rotation in binary stars}
\bigskip

\begin{center}
\author "B.~V.~Vasiliev
%\bigskip

Institute in Physical-Technical Problems, 141980, Dubna, Russia
%\bigskip

{vasiliev@dubna.ru}

\end{center}

%\bigskip
%\maketitle

\begin{abstract}
The shape of a rotating star consisting of equilibrium plasma is
considered. The velocity of apsidal rotation of close binary stars
(periastron rotation) which depends on the star shapes is
calculated. The obtained estimations are in a good agreement with
the observation data of the apsidal motion in binary systems.
\end{abstract}
%\bigskip

%\bigskip
\section{Introduction}

The apsidal motion (periastron rotation) of close binary stars is
result of a their non-Keplerian moving which originates from the
non-spherical form of stars. This non-sphericity has been produced
by a rotation of stars about their axes or by their mutual tidal
effect. The second effect is smaller usually and it can be
neglected. Following to the traditional approach to explanation of
this effect, one needs to suppose the existence of a concentration
of mass inside central part of stars. To reach an agreement
between the measuring data and calculations, it is usually
necessary to assume that the density of substance at the central
region of a star is a hundred times more than a mean density of
the star \cite{astro1}.

As it was shown earlier \cite{BV}, almost the full mass of a star
is concentrated in its plasma core at a permanent density.
Therefor the effect of periastron rotation of close binary stars
must be reviewed with the account of a change of forms of these
star cores.

According to \cite{astro1}-\cite{astro2} the ratio of the angular
velocity $\omega$ of rotation of periastron which is produced by
the rotation of a star about its axis with  the angular velocity
$\Omega$ is

\begin{equation}
\frac{\omega}{\Omega}=\frac{3}{2}\frac{(I_A-I_C)}{Ma^2}
\end{equation}
where $I_A$ and $I_C$ are the moments of inertia relatively to
principal axes of the ellipsoid. Their difference is

\begin{equation}
I_A-I_C=\frac{M}{5}(a^2-c^2),
\end{equation}
where $a$ and $c$ are the equatorial and polar radii of the star.

%It is generally accepted to consider that the angular velocity on
%the ellipse and the angular velocity of the rotation of a star
%about its axis  are coinciding and equal.
Thus we have

\begin{equation}
\frac{\omega}{\Omega}\approx \frac{3}{10}\frac{(a^2-c^2)}{a^2}.
\end{equation}

\section{The equilibrium form of the core of a rotating star}
In the absence of rotation the equilibrium equation  of plasma
inside star is \cite{BV}

\begin{equation}
\gamma {\bf g}_G+\rho_G {\bf E}_G=0\label{qm}
\end{equation}
where $\gamma$,${\bf g}_G$, $\rho_G$ and ${\bf E}_G$ are the
substance density the acceleration of gravitation, gravity-induced
density of charge and intensity of gravity-induced electric field
($div~{\bf g}_G=4\pi~ G~ \gamma$, $div~{\bf E}_G=4\pi \rho_G$ and
$\rho_G=\sqrt{G}\gamma$).

One can suppose, that at a rotation,  under action of a rotational
acceleration  ${\bf g}_\Omega$, an additional electric charge with
density $\rho_\Omega$ and electric field ${\bf E}_\Omega$ can
exist, and the equilibrium equation obtains the form:

\begin{equation}
(\gamma_G+\gamma_\Omega)({\bf g}_G+{\bf
g}_\Omega)=(\rho_G+\rho_\Omega)({\bf E}_G+{\bf E}_\Omega),
\end{equation}

where

\begin{equation}
div~({\bf E}_G+{\bf E}_\Omega)=4\pi(\rho_G+\rho_\Omega)
\end{equation}

or

\begin{equation}
div~{\bf E}_\Omega=4\pi\rho_\Omega.
\end{equation}

We can look for a solution for electric potential in the form

\begin{equation}
\varphi=C_\Omega~r^2(3cos^2\theta-1)
\end{equation}

or in Cartesian coordinates

\begin{equation}
\varphi=C_\Omega(3z^2-x^2-y^2-z^2)
\end{equation}

where $C_\Omega$ is a constant.

 Thus

\begin{equation}
E_x=2~C_\Omega~x,~ E_y=2~C_\Omega~y,~ E_z=-4~C_\Omega~z
\end{equation}

and

\begin{equation}
div~{\bf E}_\Omega=0
\end{equation}

and we obtain the important equations:

\begin{equation}
\rho_\Omega=0;
\end{equation}

\begin{equation}
\gamma g_\Omega=\rho {\bf E}_\Omega.
\end{equation}

Since a centrifugal force must be contra-balanced by the electric
force

\begin{equation}
\gamma~2\Omega^2~x=\rho~2C_\Omega~x
\end{equation}

and

\begin{equation}
C_\Omega=\frac{\gamma~\Omega^2}{\rho}=\frac{\Omega^2}{\sqrt{G}}
\end{equation}

The potential of a positively uniformly charged ball is

\begin{equation}
\varphi(r)=\frac{Q}{R}\biggl(\frac{3}{2}-\frac{r^2}{2R^2}\biggr)
\end{equation}

The negative  charge on the surface of a sphere induces inside the
sphere the potential

\begin{equation}
\varphi(R)=-\frac{Q}{R}
\end{equation}

where accordingly to Eq.({\ref{qm}}) $Q=\sqrt{G}M$, and $M$ is the
mass of the star.

Thus the total potential inside the considered star is

\begin{equation}
\varphi_\Sigma=\frac{\sqrt{G}M}{2R}\biggl(1-\frac{r^2}{R^2}\biggr)+\frac{\Omega^2}{\sqrt{G}}r^2(3cos^2\theta-1)
\end{equation}

Since the  electric potential must be equal to zero on the surface
of the star, at $r=a$ and $r=c$

\begin{equation}
\varphi_\Sigma=0
\end{equation}

and  we obtain the equation which describes the equilibrium form
of the core of a rotating star (at $\frac{a^2-c^2}{a^2}\ll 1$)

\begin{equation}
\frac{a^2-c^2}{a^2}\approx\frac{9}{2\pi}\frac{\Omega^2}{G\gamma}\label{ef}.
\end{equation}

\section{The angular velocity of the apsidal rotation}

Taking into account of Eq.({\ref{ef}}) we have

\begin{equation}
\frac{\omega}{\Omega}\approx
\frac{27}{20\pi}\frac{\Omega^2}{G\gamma}\label{oo}
\end{equation}
If both stars of a close pair induce a rotation of periastron,
this equation transforms to

\begin{equation}
\frac{\omega}{\Omega}\approx
\frac{27}{20\pi}\frac{\Omega^2}{G}\biggl(\frac{1}{\gamma_1}+\frac{1}{\gamma_2}\biggr),
\end{equation}
where $\gamma_1$ and $\gamma_2$ are densities of star cores.

The equilibrium density of star cores is known \cite{BV}:

\begin{equation}
\gamma=\frac{16}{9\pi^2}\frac{A}{Z}m_p\frac{(Z+1)^3}{a_B^3}\label{go},
\end{equation}
where $A$ and $Z$ are the mass number and charge of nuclei of
plasma, $m_p$ is proton mass, and the Borh radius is

\begin{equation}
a_B=\frac{\hbar^2}{m_e e^2}.
\end{equation}

If we introduce  the period of ellipsoidal rotation
$P=\frac{2\pi}{\Omega}$ and  the period of the rotation of
periastron $U=\frac{2\pi}{\omega}$,  we obtain from
Eq.({\ref{oo}})

\begin{equation}
\frac{P}{U}\biggl(\frac{P}{T}\biggr)^2\approx\sum_1^2\xi_i\label{2},
\end{equation}
where
\begin{equation}
T=\sqrt{\frac{243~\pi^3}{80}}~\tau_0\approx 10 \tau_0,
\end{equation}
\begin{equation}
\tau_0=\sqrt{\frac{a_B^3}{G~m_p}}\approx 7.7\cdot 10^2 sec
\end{equation}
and
\begin{equation}
\xi_i=\frac{Z_i}{A_i(Z_i+1)^3}.
\end{equation}

\section{The comparison of the calculated angular velocity of the periastron rotation with observations}

Because the substance density (Eq.({\ref{go}})) is depending
approximately on the second power of the nuclear charge, the
periastron moving of stars consisting of heavy elements will fall
out from the observation as it is very slow. Practically the
obtained equation ({\ref{2}}) shows that it is possible to observe
the periastron rotation of a star consisting of light elements
only.

The value $\xi=Z/[A(Z+1)^3]$ is equal to $1/8$ for hydrogen,
$0.0625$ for deuterium, $1.85\cdot 10^{-2}$ for helium. The
resulting value of the periastron rotation of double stars will be
the sum of separate stars rotation. The possible combinations of a
couple and their value of $\sum_1^2\xi_i$ for stars consisting of
light elements is shown in Table 1.

\bigskip

\begin{tabular}{||c|c|c||}\hline\hline
  star1&star2 &$\xi_1+\xi_2$\\
  composed of &composed of&\\\hline
  H & H & .25\\
  H & D & 0.1875\\
  H & He & 0.143\\
  H & hn & 0.125\\
  D & D & 0.125 \\
  D & He & 0.0815 \\
  D & hn & 0.0625 \\
  He & He & 0.037 \\
  He & hn & 0.0185 \\ \hline\hline
\end{tabular}
Table 1.
\bigskip

There "hn" notation in Table 1 indicates that the second component
of the couple consists of heavy elements or it is a dwarf.

The periods $U$ and $P$ are measured for few tens of close binary
stars. The data of these measurement is summarized in the Table 2.
In this table $U$ is the period of the periastron rotation in
years, $P$ is the period of the orbital rotation in astronomical
days. $M_1/M_\odot$ and $M_2/M_\odot$ are masses of the first and
the second star over the solar mass, $R_1/R_\odot$ and
$R_2/R_\odot$ are the first star radius and the second star radius
over the solar radius, $T_1$ and $T_2$ are the surface
temperatures  of the first and the second star, $a/R_\odot$ is the
orbital radius of the couple over solar radius. All these data and
references was given to us by Dr.Khaliullin K.F. (Sternberg
Astronomical Institute) \cite{Kh}.

\begin{figure}
\begin{center}
\includegraphics[8cm,2cm][14cm,12cm]{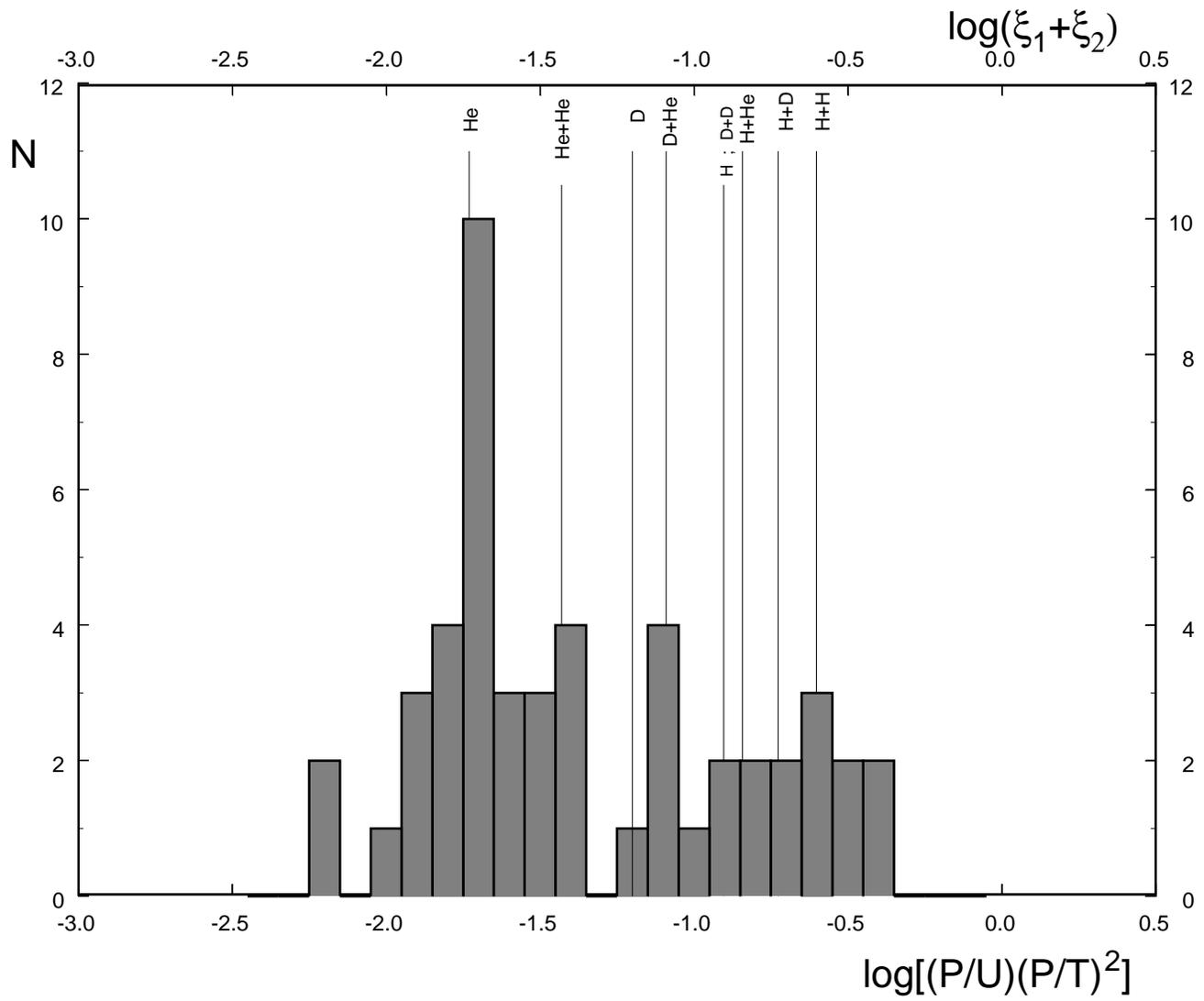}
%\vspace{11cm}
\caption {The distribution of binary stars on value of
$(P/U)(P/T)^2$.} \label{periastr}
\end{center}
\end{figure}

One can compare our calculation with the data of these
measurements. The distribution of close binary stars on value of
$(P/U) (P/T)^2$ is shown on Fig.{\ref{periastr}} in logarithmic
scale. The lines mark the values of parameters $\sum_1^2\xi_i$ for
different pairs of binary stars. It can be seen that the
calculated values the periastron rotation for stars composed by
light elements which is summarized in Table 1 are in the good
agreement with separate peaks of measured data. It confirms that
our approach to interpretation of this effect and is adequate to
produce  the satisfactory accuracy of estimations.

Author expresses sincere thanks to Dr.Khaliullin K.F. (Sternberg
Astronomical Institute) for the affording  the opportunity to
compare obtained estimations with the measured data.

%\clearpage

%\vspace{1cm}
\hspace{-2cm}
%\begin{center}
{\scriptsize
\begin{tabular}
{||r|l|r|r|r|r|r|r|r|r|r|c||}\hline\hline
%\begin{tabular}{||c|c|c|c|c|c|c|c|c|c|c|c||}\hline\hline
  N & Name of star & U & P & $M_1/M_{\odot}$ & $M_2/M_{\odot}$ & $a/R_{\odot}$ & $R_1/R_{\odot}$ & $R_2/R_{\odot}$& $T_1$ & $T_2$ \\\hline
  1 & BW Aqr & 5140 & 6.720 & 1.48 & 1.38 & 21.26 & 1.803 & 2.075 & 6100 & 6000 \\
  2 & V 889 Aql & 23200 & 11.121 & 2.40 & 2.20& 34.85 & 2.028 & 1.826 & 9900 & 9400  \\
  3 & V 539 Ara & 150 & 3.169 & 6.24 & 5.31 & 20.51 & 4.512 & 3.425 & 17800 & 17000   \\
  4 & AS Cam & 2250 & 3.431 & 3.31 & 2.51 & 17.21 & 2.580 & 1.912 & 11500 & 10000  \\
  5 & EM Car & 42 & 3.414 & 22.80 & 21.40 & 33.74 & 9.350 & 8.348 & 33100 & 32400 \\
  6 & GL Car & 25 & 2.422 & 13.50 & 13.00 & 13.28 & 4.998 & 4.726 & 28800 & 28800  \\
  7 & QX Car & 361 & 4.478 & 9.27 & 8.48 & 29.81 & 4.292 & 4.054 & 23400 & 22400  \\
  8 & AR Cas & 922 & 6.066 & 6.70 & 1.90 & 28.66 & 4.591 & 1.808 & 18200 & 8700  \\
  9 & IT Cas & 404 & 3.897 & 1.40 & 1.40 & 14.68 & 1.616 & 1.644 & 6450 & 6400 \\
  10 & OX Cas & 40 & 2.489 & 7.20 & 6.30 & 18.30 & 4.690 & 4.543 & 23800 & 23000  \\
  11 & PV Cas & 91 & 1.750 & 2.79 & 2.79 & 10.83 & 2.264 & 2.264 & 11200 & 11200  \\
  12 & KT Cen & 260 & 4.130 & 5.30 & 5.00 & 23.56 & 4.028 & 3.745 & 16200 & 15800 \\
  13 & V 346 Cen & 321 & 6.322 & 11.80 & 8.40 & 39.16 & 8.263 & 4.190 & 23700 & 22400  \\
  14 & CW Cep & 45 & 2.729  & 11.60 & 11.10 & 23.32 & 5.392 & 4.954 & 26300 & 25700  \\
  15 & EK Cep & 4300 & 4.428 & 2.02 & 1.12 & 16.61 & 1.574 & 1.332 & 10000  & 6400  \\
  16 & $\alpha$ Cr B & 46000 & 17.360 & 2.58 & 0.92 & 42.81 & 3.314 & 0.955 & 9100 & 5400  \\
  17 & Y Cyg & 48 & 2.997 & 17.50 & 17.30 & 28.54 & 6.022 & 5.680 & 33100 & 32400 \\
  18 & Y 380 Cyg & 1550 & 12.426 & 14.30 & 8.00 & 63.51 & 17.080 & 4.300 & 20700 & 21600  \\
  19 & V 453 Cyg & 71 & 3.890 & 14.50 & 11.30 & 30.74 & 8.607 & 5.410 & 26600 & 26000 \\
  20 & V 477 Cyg & 351 & 2.347 & 1.79 & 1.35 & 10.88 & 1.567 & 1.269 & 8550 & 6500 \\
  21 & V 478 Cyg & 26 & 2.881 & 16.30 & 16.60 & 27.29 & 7.422 & 7.422 & 29800 & 29800  \\
  22 & V 541 Cyg & 40000 & 15.338 & 2.69 & 2.60 & 45.24 & 2.013 & 1.900 & 10900 & 10800 \\
  23 & V 1143 Cyg & 10300 & 7.641 & 1.39 & 1.35 & 22.83 & 1.440 & 1.226 & 6500 & 6400 \\
  24 & V 1765 Cyg & 1932 & 13.374 & 23.50 & 11.70 & 77.64 & 19.960 & 6.522 & 25700 & 25100  \\
  25 & DI Her & 29000 & 10.550  & 5.15 & 4.52 & 43.10 & 2.478 & 2.689 & 17000 & 15100 \\
  26 & HS Her & 92 & 1.637  & 4.25 & 1.49 & 10.46 & 2.709 & 1.485 & 15300 & 7700 \\
  27 & CO Lac & 44 & 1.542 & 3.13 & 2.75 & 10.13 & 2.533 & 2.128 & 11400 & 10900 \\
  28 & GG Lup & 101 & 1.850 & 4.12 & 2.51 & 13.22 & 2.644 & 1.917 & 14400 & 10500 \\
  29 & RU Mon & 348 & 3.585 & 3.60 & 3.33 & 18.78 & 2.554 & 2.291 & 12900 & 12600  \\
  30 & GN Nor & 500 & 5.703 & 2.50 & 2.50 & 22.96 & 4.591 & 4.591 & 7800 & 7800  \\
  31 & U Oph & 21 & 1.677 & 5.02 & 4.52 & 12.59 & 3.311 & 3.110 & 16400 & 15200 \\
  32 & V 451 Oph & 170 & 2.197 & 2.77 & 2.35 & 12.25 & 2.538 & 1.862 & 10900 & 9800  \\
  33 & $\beta$ Ori & 228 & 5.732 & 19.80 & 7.50 & 40.56 & 14.160 & 8.072 & 26600 & 17800  \\
  34 & FT Ori & 481 & 3.150 & 2.50 & 2.30 & 15.24 & 1.890 & 1.799 & 10600 & 9500 \\
  35 & AG Per & 76 & 2.029 & 5.36 & 4.90 & 14.65 & 2.995 & 2.606 & 17000 & 17000 \\
  36 & IQ Per & 119 & 1.744 & 3.51 & 1.73 & 10.58 & 2.445 & 1.503 & 13300 & 8100 \\
  37 & $\zeta$ Phe & 44 & 1.670 & 3.93 & 2.55 & 11.04 & 2.851 & 1.852 & 14100 & 10500  \\
  38 & KX Pup & 170 & 2.147 & 2.50 & 1.80 & 11.38 & 2.333 & 1.593 & 10200 & 8100 \\
  39 & NO Pup & 37& 1.257 & 2.88 & 1.50 & 8.01 & 2.028 & 1.419 & 11400 & 7000 \\
  40 & VV Pyx & 3200 & 4.596 & 2.10 & 2.10 & 18.76 & 2.167 & 2.167 & 8700 & 8700  \\
  41 & YY Sgr & 297 & 2.628 & 2.36 & 2.29 & 13.37 & 2.196 & 1.992 & 9300 & 9300  \\
  42 & V 523 Sgr & 203 & 2.324 & 2.10 & 1.90 & 11.71 & 2.682 & 1.839 & 8300 & 8300  \\
  43 & V 526 Sgr & 156 & 1.919 & 2.11 & 1.66 & 10.11 & 1.900 & 1.597 & 7600 & 7600 \\
  44 & V 1647 Sgr & 592 & 3.283 & 2.19 & 1.97 & 14.94 & 1.832 & 1.669 & 8900 & 8900  \\
  45 & V 2283 Sgr & 570 & 3.471 & 3.00 & 2.22 & 16.72 & 1.957 & 1.656 & 9800 & 9800  \\
  46 & V 760 Sco & 40 & 1.731 & 4.98 & 4.62 & 12.89 & 3.015 & 2.642 & 15800 & 15800 \\
  47 & AO Vel & 50 & 1.585 & 3.20 & 2.90 & 11.41 & 2.623 & 2.954 & 10700 & 10700 \\
  48 & EO Vel & 1600 & 5.330 & 3.21 & 2.77 & 23.29 & 3.145 & 3.284 & 10100 & 10100  \\
  49 & $\alpha$ Vir & 140 & 4.015 & 10.80 & 6.80 & 27.64 & 8.097 & 4.394 & 19000 & 19000 \\
  50 & DR Vul & 36 & 2.251 & 13.20 & 12.10 & 21.21 & 4.814 & 4.369 & 28000 & 28000 \\ \hline\hline
\end{tabular}}
%\end{center}


\begin{thebibliography}{2}


\bibitem {astro1} Russel H.N., Monthly Notices of the RAS {\bf{88}} (1928) 642

\bibitem {astro2} Chandrasekhar S., Monthly Notices of the RAS {\bf{93}} (1933) 449


\bibitem {BV}   Vasiliev B.V. - Nuovo Cimento B, 2001, v.116, pp.617-634.


\bibitem {Kh}    Khaliullin K.F., private comunication.


\end{thebibliography}
\end{document}